\documentstyle[epsfig]{mn}

\newcommand{\mnras}{MNRAS}
\newcommand{\apj}{ApJ}

\title[]
{Assembly bias in the clustering of dark matter haloes}
\author[L. Gao \& S.~D.~M. White]  {Liang ~Gao $^1$\thanks{Email:
        liang.gao@durham.ac.uk}, Simon D.~M.~White$^2$ \\
      $^1$ Department of Physics, University of Durham, South
      Road, Durham, DH1 3LE\\
      $^2$  Max--Planck--Institut f\"ur Astrophysik, D-85748 Garching,
        Germany}
\begin{document}
\label{firstpage} \maketitle

\begin{abstract}
We use a very large simulation of structure growth in a $\Lambda$CDM universe
-- the Millennium Simulation -- to study assembly bias, the fact that the
large-scale clustering of haloes of given mass varies significantly with their
assembly history. We extend earlier work based on the same simulation by
superposing results for redshifts from 0 to 3, by defining a less noisy
estimator of clustering amplitude, and by considering halo concentration,
substructure mass fraction and spin, as well as formation time, as additional
parameters. These improvements lead to results with less noise than previous
studies and covering a wider range of halo masses and structural properties.
We find significant and significantly different assembly bias effects for all
the halo properties we consider, although in all cases the dependences on halo
mass and on redshift are adequately described as a dependence on equivalent
peak height $\nu(M,z)$. The $\nu$-dependences for different halo properties
differ qualitatively and are not related as might naively be expected given
the relations between formation time, concentration, substructure fraction and
spin found for the halo population as a whole. These results suggest that it
will be difficult to build models for the galaxy populations of dark haloes
which can robustly relate the amplitude of large-scale galaxy clustering
to that for mass clustering at better than the 10\% level.
\end{abstract}

\title{Halo assembly bias}

\section{Introduction}

Gao, Springel \& White (2005) used the very large Millennium Simulation of
Springel et al. (2005) to demonstrate that the standard $\Lambda$CDM paradigm
predicts the clustering of dark matter haloes to depend not only on their mass
but also on their formation time. The effect is strong for low-mass haloes but
weak or absent at high mass. This result is surprising since it is
incompatible with the standard excursion set theory for the growth of
structure (e.g. Bond et al. 1991; Lacey \& Cole 1993, Mo \& White 1996), and
it contradicts a fundamental assumption of the halo occupation distribution
(HOD) models often used to study galaxy clustering, namely that the galaxy
content of a halo of given mass is statistically independent of its larger
scale environment (e.g. Kauffmann, Nusser \& Steinmetz 1997; Jing, Mo
\& Boerner 1998; Peacock \& Smith 2000; Benson et al. 2000; Berlind et
al. 2003; Yang, Mo \& van den Bosch 2003).

Subsequent studies confirmed this result (Harker et al. 2006; Zhu et al. 2006;
Wechsler et al. 2006; Wetzel et al. 2007; Jing, Suto \& Mo 2007). In addition,
Wechsler et al. (2006) demonstrated a dependence of halo clustering on halo
concentration, noting it to have similar strength at high and low mass, but
opposite sign; concentrated haloes are the most clustered at low mass, but the
least clustered at high mass. This is also surprising since there is a tight
correlation between halo formation time and halo concentration (e.g. Navarro,
Frenk \& White 1997; Wechsler et al. 2001; Zhao et al. 2003) so one might
expect a similar dependence on the two properties.  More recently, Bett et
al. (2007) studied clustering as a function of halo spin and shape, finding
stronger clustering at given mass for larger spin and for smaller
major-to-minor axis ratio. (See also Hahn et al. (2007) for a slightly
different approach to studying environmental effects on halo spin.)
Interestingly, the dependences on shape and spin are stronger for high-mass
haloes.

We use the term ``assembly bias'' to describe all such dependences, since they
show that the spatial distribution of haloes depends not only on their mass
but also on the details of their assembly history. Such details are
undoubtedly reflected in the galaxy populations they host, so one may expect
assembly bias to affect galaxy clustering in a way which is not easily
represented in a simple HOD model. A first assessment of the strength of such
effects was made by Croton, Gao \& White (2006) using a direct simulation of
galaxy formation within the Millennium Simulation, while possible
observational evidence for them has been cited by Yang, Mo \& van den Bosch
(2006), Blanton, Berlind \& Hogg (2007) and Berlind et al. (2007).  Some
recent theoretical work has explored extensions of excursion set theory which
may allow a description of halo assembly bias (Wang, Mo \& Jing 2007, Sandvik
et al. 2007, Zentner 2007) but it remains unclear whether these approaches can
account for the results discussed above and in this paper.

In this Letter we extend the work of Gao, Springel \& White (2005), again
using the Millennium Simulation, by defining a less noisy estimator of
clustering strength, by combining results for a number of redshifts, and by
considering clustering as a function of halo properties other than formation
time. As a result, we can study assembly bias in more detail and over a
substantially wider halo mass range than in the earlier paper. In Section~2 we
summarise the relevant properties of the simulation and of the halo database
that we study. In Section~3, we explore halo assembly bias as a function of
halo mass and of a variety of halo structural properties. Finally we give a
short summary and discussion.

\section{The simulation}
The Millennium Simulation was carried out by the Virgo Consortium in 2004
on an IBM Regatta system at the Max Planck Society's supercomputer centre in
Garching (Springel et al.  2005).  It adopted concordance values for the
parameters of a flat $\Lambda$CDM cosmological model, $\Omega_{\rm dm}=0.205$,
$\Omega_{\rm b}=0.045$ for the current densities in Cold Dark Matter and
baryons, $h=0.73$ for the present dimensionless value of the Hubble constant,
$\sigma_8=0.9$ for the {\it rms} linear mass fluctuation in a sphere of radius
$8 h^{-1}$Mpc extrapolated to $z=0$, and $n=1$ for the slope of the primordial
fluctuation spectrum. The simulation followed $2160^3$ dark matter particles
from $z=127$ to the present-day within a cubic region $500 h^{-1}$Mpc on a
side. The individual particle mass is thus $8.6\times 10^{8}h^{-1}{\rm
M_\odot}$. The gravitational force had a Plummer-equivalent comoving softening
of $5h^{-1}$kpc. Initial conditions were set using the Boltzmann code {\small
CMBFAST} (Seljak \& Zaldarriaga 1996) to generate a realisation of the desired
power spectrum which was then imposed on a glass-like uniform particle load
(White 1996).

The {\small TREE-PM} N-body code {\small GADGET2} (Springel 2005)
was used to carry out the simulation and the full data were stored
at $64$ times spaced approximately equally in the logarithm of the
expansion factor at early times and at approximately 200 Myr
intervals after $z=1$. At run time all collapsed haloes with at
least 20 particles were identified using a friends-of-friends
({\small FOF}) group-finder with linking parameter $b=0.2$ (Davis
et al. 1985). Post-processing with the substructure algorithm
{\small SUBFIND} (Springel et al. 2001) allowed a variety of
internal structural properties to be measured for all these haloes
and for their resolved subhaloes.  This in turn allowed trees to
be built which store detailed assembly histories for every object
and its substructure. In this paper we concentrate on haloes with
{\ small FOF} mass corresponding to 65 particles or more. At
redshifts 3, 2, 1 and 0 there are $4.6\times 10^6$, $5.8\times
10^6$, $6.2\times 10^6$ and $5.7\times 10^6$ such {\small FOF}
haloes, respectively. The halo/subhalo data and their associated
structure formation trees are publically available (together with
galaxy data from two independent galaxy formation models) through
a database at http://www.mpa-garching.mpg.de/millennium

\section{Halo assembly bias}
\begin{figure*}
\vspace{-0.6cm}
\resizebox{7cm}{!}{\includegraphics{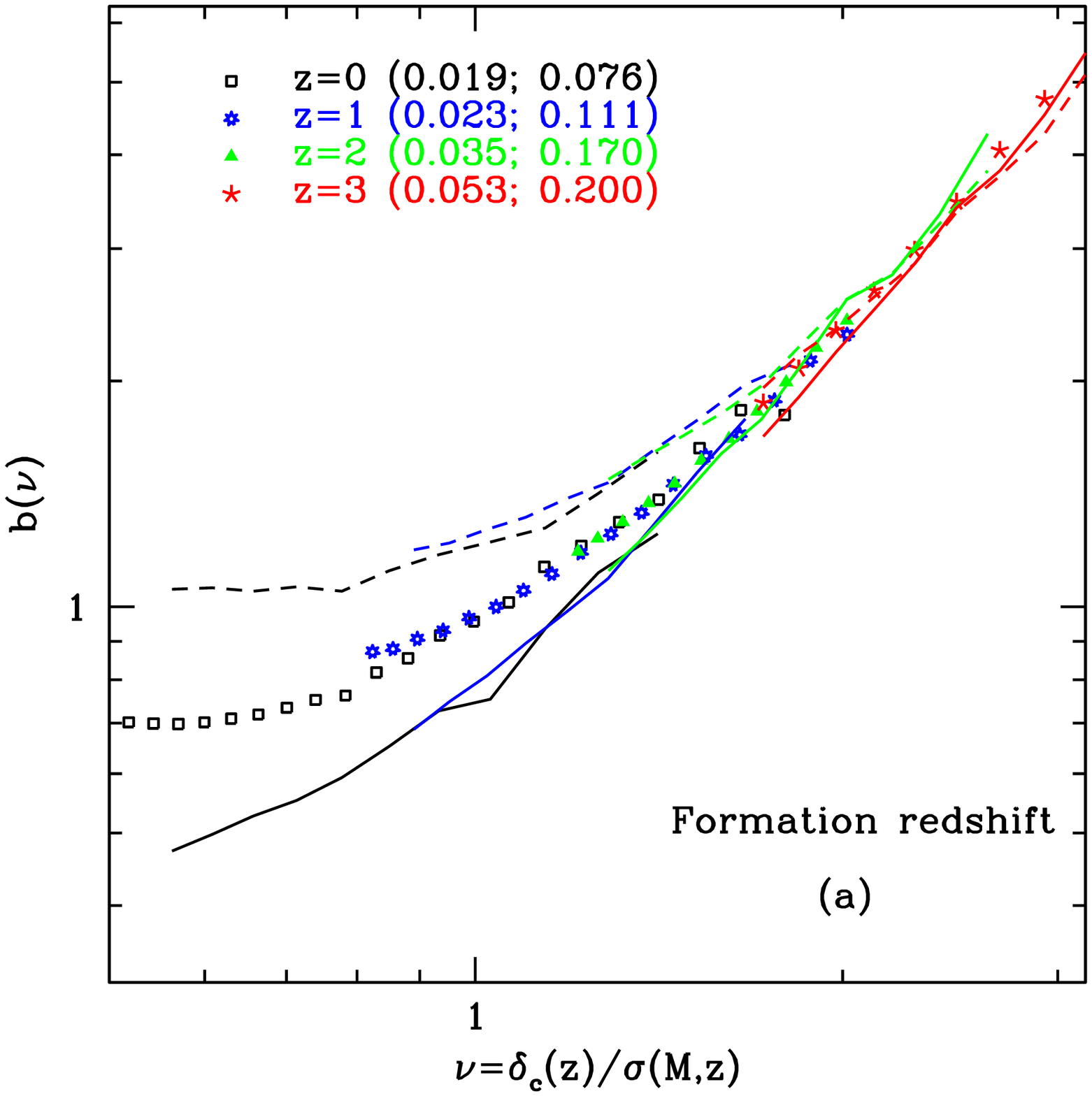}}
\hspace{-0.23cm}\resizebox{7cm}{!}{\includegraphics{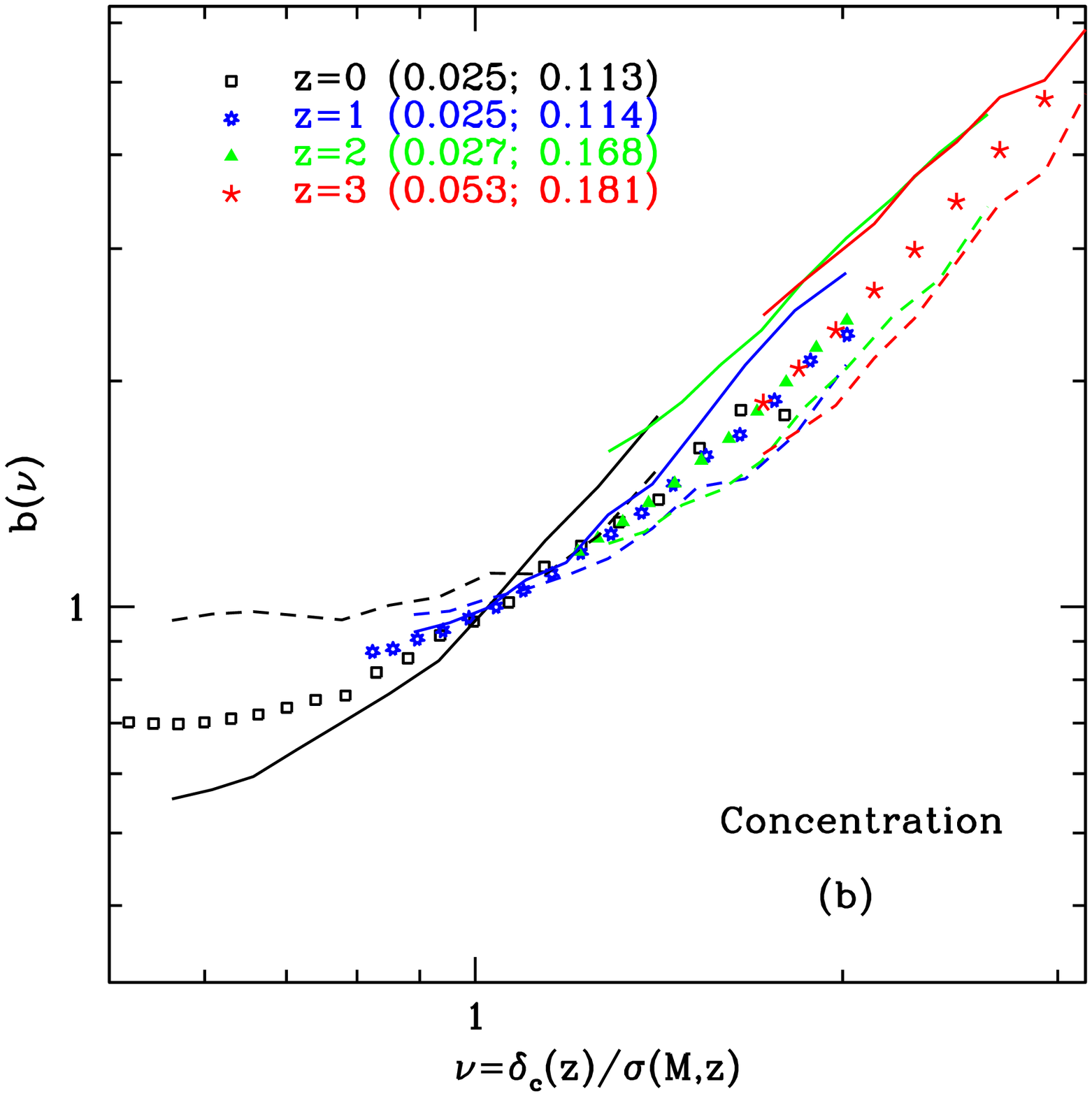}}\\
\vspace{-0.7cm} \hspace{-0.23cm}
\resizebox{7cm}{!}{\includegraphics{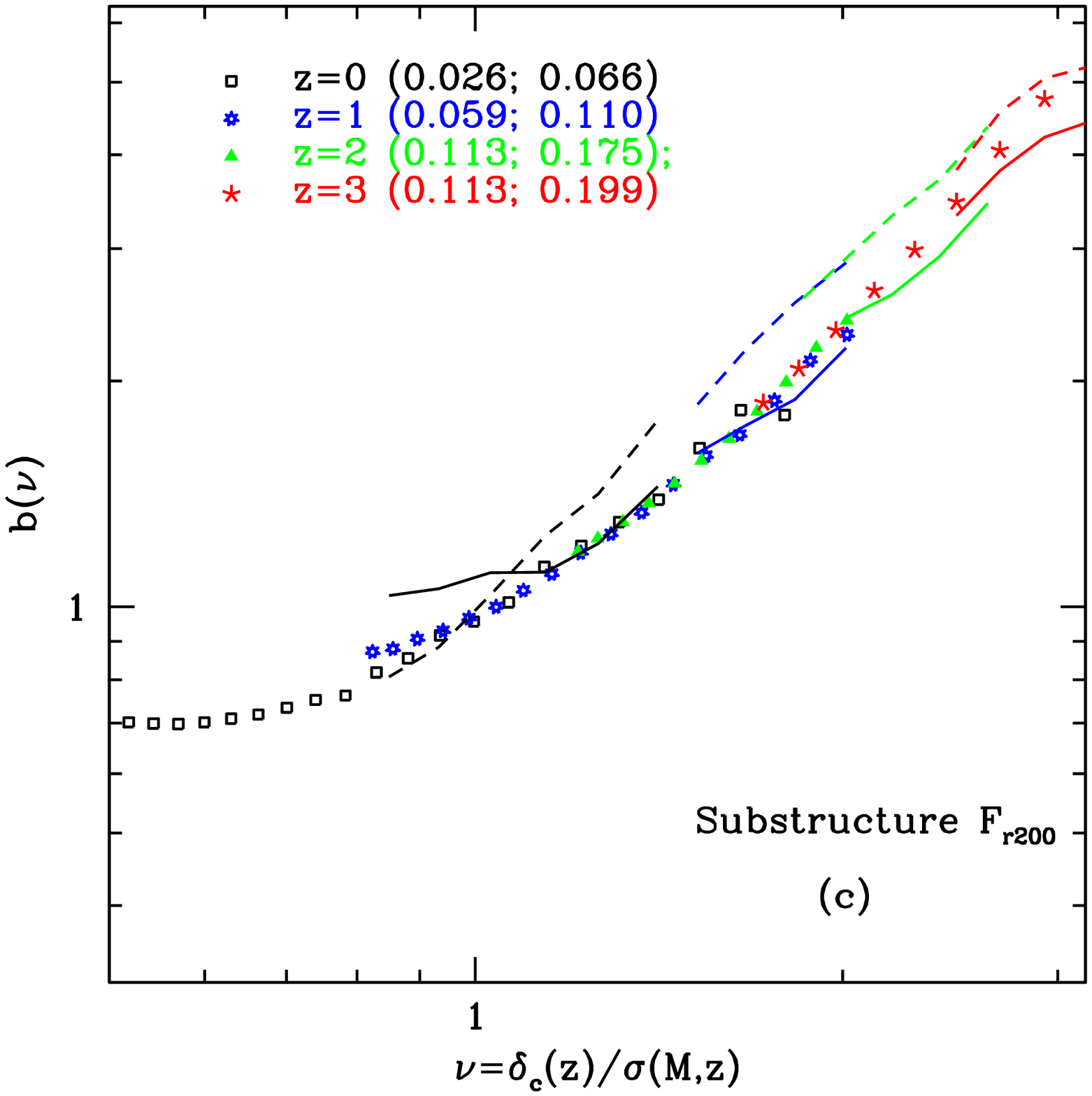}}
\hspace{-0.23cm}\resizebox{7cm}{!}{\includegraphics{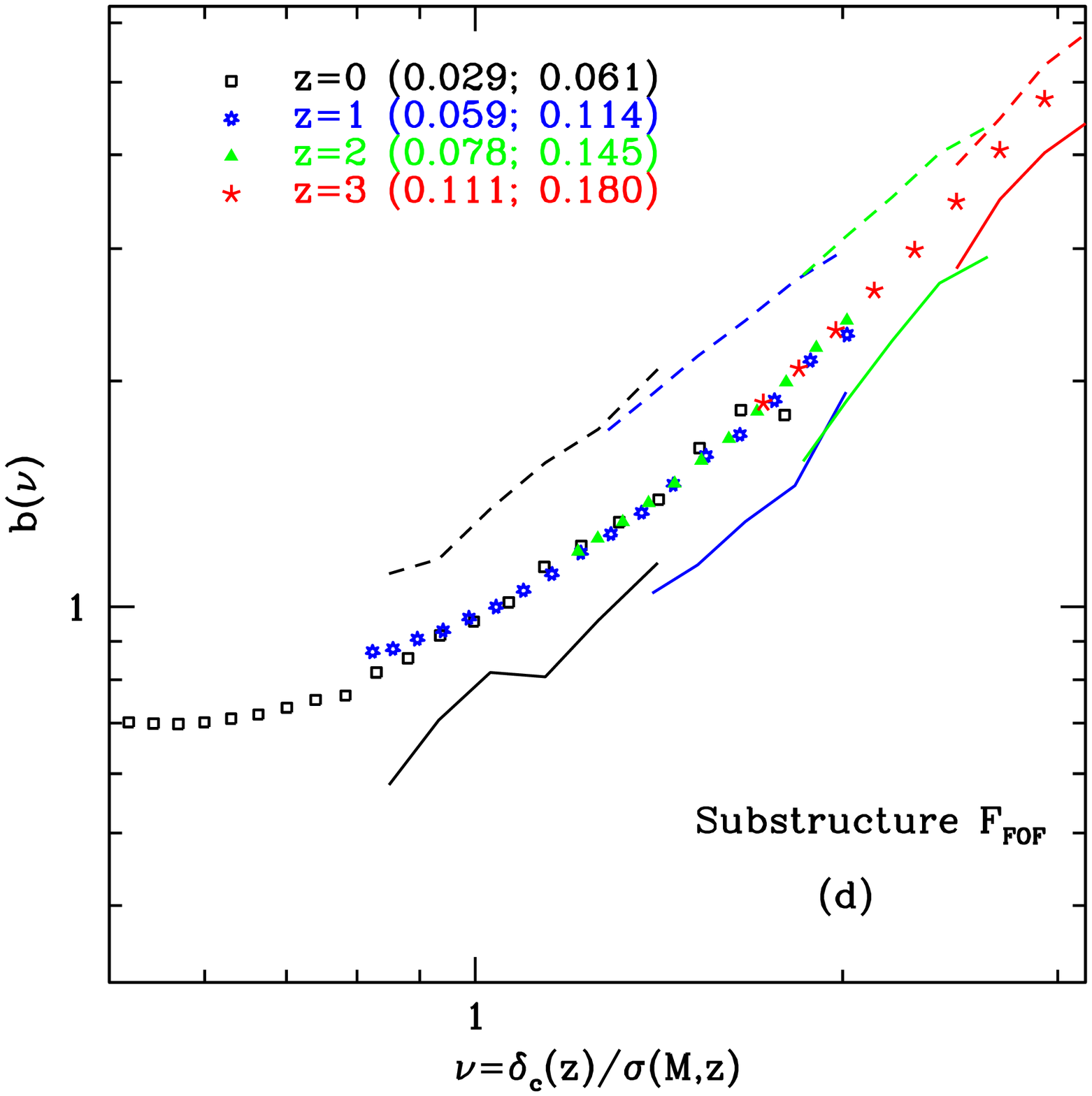}}
\vspace{-0.7cm}
\hspace{-0.23cm}\resizebox{7cm}{!}{\includegraphics{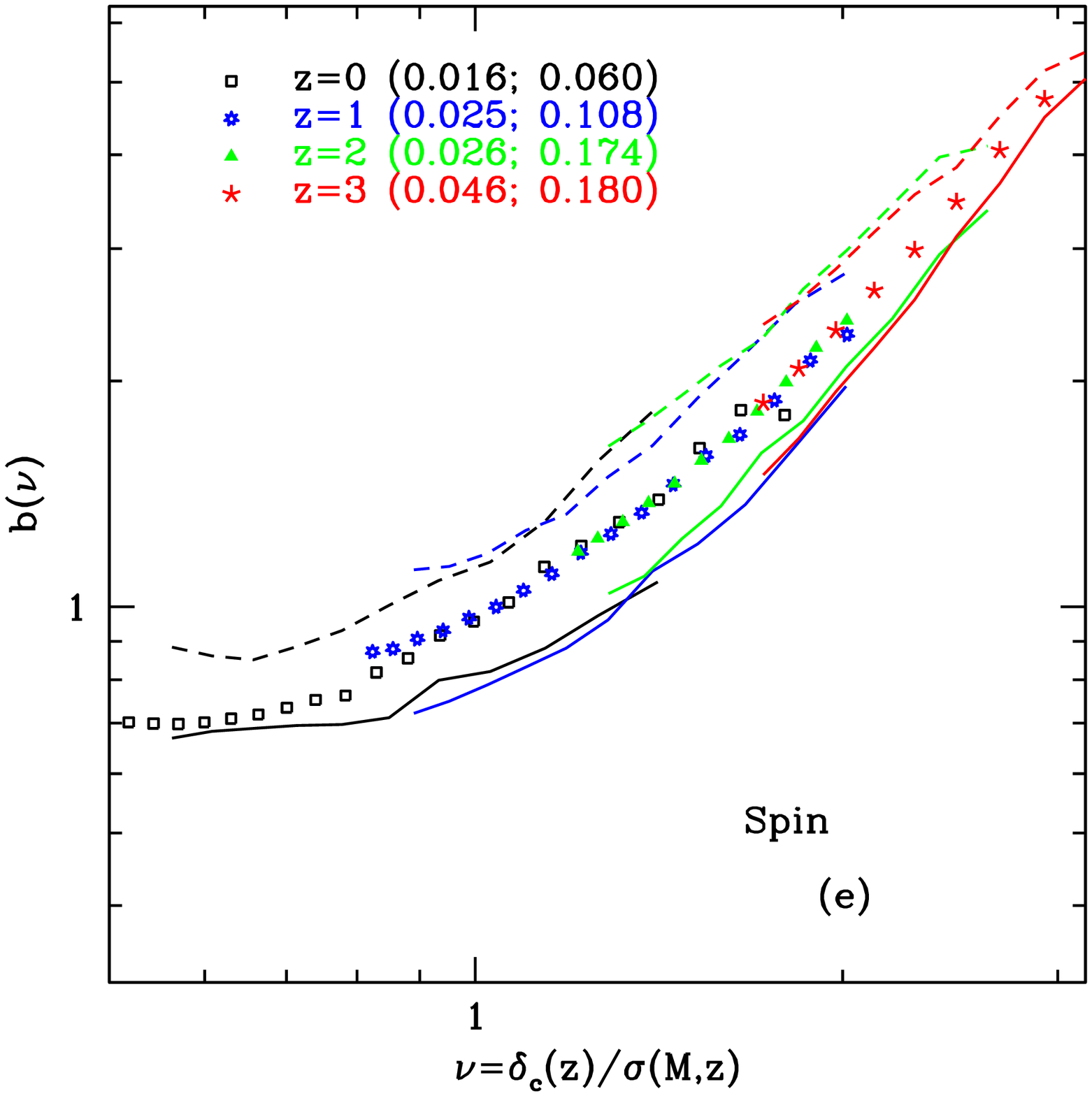}}
\hspace{-0.23cm}\resizebox{7cm}{!}{\includegraphics{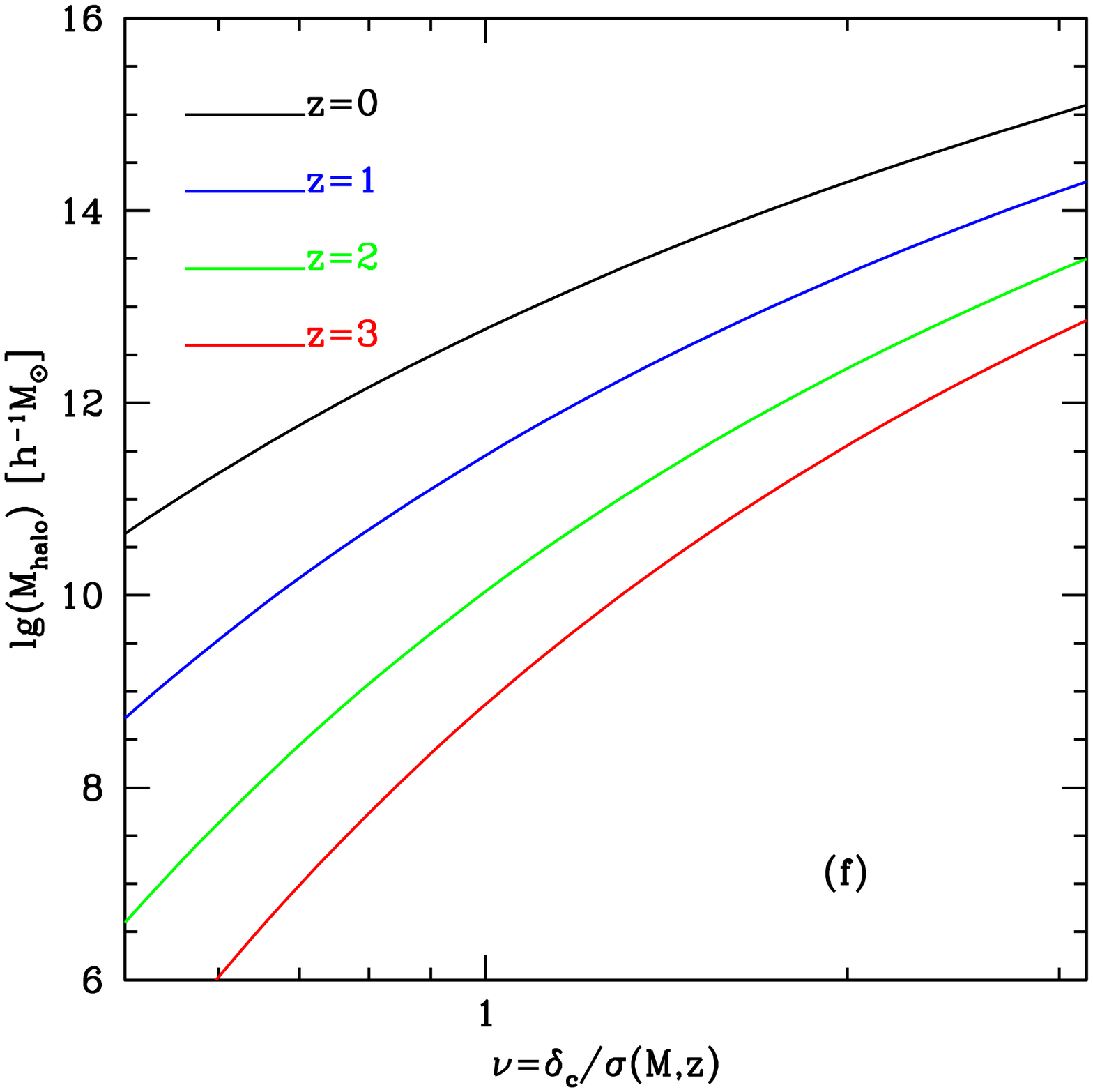}}
\caption{Bias factor as a function of halo mass and halo properties. Halo mass
  is given parametrically through the equivalent peak height $\nu(M,z) =
  \delta(M)/\delta_c(z)$. The additional properties in the first five panels
  are: (a) formation redshift, (b) concentration, (c) substructure mass
  fraction within $r_{200}$, (d) mass fraction in the main subhalo of the
  {\small FOF} halo, and (e) spin. The symbols repeated in all five panels are
  bias factors defined for all haloes of the relevant mass. The solid and
  dashed curves in these panels are bias factors for haloes in the lower and
  upper 20\% tails of the distributions of the particular property
  concerned. Panel (f) plots halo mass as a function of $\nu$ at the four
  redshifts we combined to make these plots. Both the lines in this panel and
  the symbols in the earlier panels are colour-coded according to redshift, as
  indicated by the labels. The numbers in parentheses following the redshift
  labels in each panel give the {\it rms} scatter in $b$ among 100 bootstrap
  resamplings of the halo subsamples used to obtain $b$-values for the
  solid and dashed curves at the central and at the largest $\nu$-values
  plotted for each redshift.}
  \label{fig:fig1}
\end{figure*}

In the one-dimensional excursion set model for structure formation
(e.g. Bond et al. 1991; Lacey \& Cole 1993), the formation history
of a halo is encoded in the random walk at higher mass resolution
than that which defines the halo itself, and is thus statistically
independent of the environment, which is encoded in the random
walk at lower resolution (White 1996).  This is inconsistent with
the dependences on formation time and concentration found in
simulations (Gao et al. 2005; Wechsler et al. 2006).  Here we
extend the work of Gao et al. (2005) by defining a less noisy
measure of clustering strength, by analysing data at $z=1$, $2$
and $3$ as well as $z=0$, and by examining assembly bias as a
function of properties other than formation time.

The specific halo properties which we will consider in this Letter are:
\begin{description}
\item[(1)] {\bf Formation Time}~~~~Following Gao et al. (2004, 2005), we
define the formation time of a dark halo as the redshift when half of its
final mass was first assembled into a single object.  We use the stored
merging tree to find the earliest time when its most massive progenitor had
more than half the final mass, then interpolate linearly between this and the
immediately preceding output to estimate the redshift when this progenitor
had exactly half the final mass. This we define as the formation time.

\item[(2)] {\bf Concentration}~~~~We characterise the concentration of a dark
halo by the ratio $V_{max}/V_{200}$ of the peak value of its circular velocity
curve to the circular velocity at $r_{200}$. Here $V_c(r) = (GM(r)/r)^{1/2}$
and $V_{200} = V_c(r_{200})$, where $r_{200}$ is the radius enclosing a mean
overdensity 200 times the critical value. $V_{max}$ is the maximum value
of $V_c$ for $r<r_{200}$, evaluated using only particles bound to the main
subhalo of the {\small FOF} halo. This definition has the advantages of being
robust and of not requiring any model to be fit to the simulation data.

\item[(3)] {\bf Subhalo mass fraction within $r_{200}$}~~~~As one measure of
the amount substructure, we consider the fraction $F_{r200}$ of the total mass
within $r_{200}$ in the form of self-bound substructures detected by {\small
SUBFIND}. Note that we exclude the main subhalo, which we identify with the
body of the halo itself, and that {\small SUBFIND} only catalogues subhaloes
made up of 20 or more particles. In practice, we find that this measure can
only be used effectively to rank haloes with {\small FOF} masses above 5000
particles.  We will not consider lower mass haloes when studying clustering as
a function of $F_{r200}$.

\item{(4)} {\bf Mass fraction in the main subhalo}~~~~As an alternative
measure of the amount of substructure in a halo, we use the ratio $F_{FOF}$ of
the mass of the main self-bound subhalo to the mass of the original {\small
FOF} halo. This definition contrasts with the previous one in being sensitive
to neighboring structures beyond $r_{200}$ which are ``fortuitously'' joined
to the main halo by the {\small FOF} linking procedure. With this measure we
can study clustering as a function of substructure fraction to a {\small FOF}
mass limit of 2000 particles.

\item[(5)] {\bf Halo spin}~~~~Halo spin can be conveniently defined through
$\lambda=|\vec{J}|/(\sqrt{2}MVr_{200})$ where angular momentum $\vec{J}$ and
mass $M$ are the values within a sphere of radius $r_{200}$ and the circular
velocity $V$ is also evaluated at this radius (Bullock et al. 2001).
\end{description}

As in Gao et al. (2005), we examine assembly bias by comparing the spatial
clustering of a subset of haloes of given mass to that of the set as a whole.
In the earlier paper we estimated a bias factor $b$ for each subset of haloes
as the square root of the separation-averaged ratio of its spatial
autocorrelation function to that of the dark matter. This estimator becomes
quite noisy when the number of haloes in the subset is small. In this paper we
prefer to estimate $b$ as the ratio of the halo-mass cross-correlation to the
mass autocorrelation. Specifically, we estimate $b$ as the relative
normalisation factor which minimizes the mean square of the difference $\log
\xi_{hm} - \log b\xi_{mm}$ for four equal width bins in $\log r$ spanning the
comoving separation range $6h^{-1}{\rm Mpc} < r < 20h^{-1}{\rm Mpc}$. This
estimator has improved noise characteristics because of the large number of
dark matter particles available. According to standard halo bias models it
should be equivalent to the earlier estimator (e.g. Mo \& White 1996) and we
have verified that the two give consistent $b$ values for our halo data. In
these models the large-scale bias depends on mass and redshift through the
equivalent ``peak height''
\begin{equation}
\nu(M,z) = \delta_c(z)/\sigma(M),
\end{equation}
where $\sigma(M)$ is the rms {\it linear} overdensity (extrapolated to $z=0$)
within a sphere which in the mean contains mass $M$, and $\delta_c(z)$ is the
linear overdensity threshold for collapse at redshift $z$, again extrapolated
to the corresponding value at $z=0$.  $\sigma(M)$ depends only on the power
spectrum of initial density fluctuations, while $\delta_c(z)$ depends on the
current densities in gravitating matter and dark energy (e.g. Eke, Cole \&
Frenk 1996). Gao et al. (2005) showed that halo clustering in the Millennium
Simulation obeys this scaling over the redshift range $0\leq z \leq 5$. We
will use it to superpose results from different redshifts.

Halo assembly bias is shown in Fig. ~\ref{fig:fig1} as a function of
$\nu(M,z)$ and of the various halo properties discussed above. These plots
combine results for redshifts 0, 1, 2 and 3. In each of the first five panels
we repeat bias values for all haloes of a given mass from Figure 1 of Gao
et al. (2005). The haloes in each $\nu$ bin are ranked in terms of each of the
five properties in turn, and bias values are then estimated using our
cross-correlation method for subsets consisting of haloes in the upper and
lower 20\% tails of the distribution in this additional property. Dashed lines
show bias values for the haloes with the highest formation redshifts, the
highest concentrations, the most substructure and the most spin. Solid lines
give bias values for the opposite tails. Colours (both for symbols and for
lines) denote the redshift of the output from which the data were taken, as
indicated by the labels. Note the good agreement between results for different
redshifts where these overlap. Finally, the last panel of Fig. ~\ref{fig:fig1}
gives halo mass as a function of $\nu$ for the four redshifts used to make
this figure.  With these curves one can convert the $x$-axes in the other
panels to halo mass for $0\leq z \leq 3$.

The results for formation time in Fig. ~\ref{fig:fig1} confirm those of Gao et
al (2005) and extend them to considerably higher $\nu$. Assembly bias is
strong only for low-mass haloes, with the oldest haloes being the most
clustered. It is weak or absent at the highest masses. The largest $\nu$
values correspond to masses above $10^{15}h^{-1}M_\odot$ at $z=0$. There is a
hint that the dependence may reverse at these masses, with young haloes being
more clustered than old ones as argued by Jing, Suto \& Mo (2007), but any
such reversal is clearly very weak. Their suggestion that reversal occurs near
$\nu\sim 1.7$ is clearly not supported by our data.

The dependence of assembly bias on concentration differs qualitatively from
that on formation time.  The most concentrated haloes are the most clustered
for $\nu < 1$, but they are the least clustered for $\nu > 1$.  We have
checked that at each $\nu$ our halo samples obey the relation between
concentration and formation time first pointed out by Navarro et al. (1997);
concentrated haloes of a given mass typically formed earlier than less
concentrated haloes. Nevertheless, there is clearly a range of
$\nu$ (roughly $1<\nu<2$) where clustering is significantly stronger for older
yet also for {\it less} concentrated haloes. Overall, our results for
concentration agree well with those of Wechsler et al. (2006) and Jing, Suto
\& Mo (2007), though they are less noisy because of the better statistics
provided by the Millennium Simulation.

The dependence of assembly bias on substructure is different again in shape,
and moreover differs between our two substructure measures. For both, the
dependence varies only slowly with $\nu$, and haloes with more substructure
are almost always the more strongly clustered.  The dependence is stronger,
however, when substructure is measured by the fraction of FOF mass in the main
subhalo ($F_{FOF}$) than when it is measured by the subhalo mass fraction
within $r_{200}$ ($F_{r200}$), and it gets weaker with increasing $\nu$ in the
first case, while it strengthens in the second.  Indeed, the dependence
appears to reverse at $\nu < 1$ in the $F_{r200}$ case, although this needs to
be confirmed by a simulation of higher resolution. Note that at all $\nu$ our
halo samples obey the kind of correlation of substructure with formation time
and concentration pointed out out by Gao et al. (2004). Haloes with more
substructure tend to be younger and to have lower concentrations. Thus
the dependence of assembly bias on substructure is the opposite of
what might naively have been inferred from its dependence on formation
time, and also differs qualitatively from that on concentration.

Assembly bias also depends on halo spin; rapidly rotating haloes cluster more
strongly than slowly rotating ones, with little dependence on $\nu$.  This
agrees reasonably well with the results of Bett et al. (2007) who find a
somewhat stronger trend with halo mass. This may reflect their slightly
different (and cleaner) definition of halo spin, their different (and noisier)
estimator of bias, or the fact that they only used numerical data for $z=0$.

These plots show clearly that assembly bias depends on the physical properties
of haloes in a complex way. Typical effects are at the 10 to 30\% level in $b$
(20 to 70\% in correlation or power spectrum amplitude) and depend on several
additional parameters at fixed halo mass. Our results are, however, consistent
with the hypothesis that dependences on mass and redshift can be combined into
a dependence on the single parameter $\nu$.

Finally we note that because of the large size of the Millennium
Simulation the formal errors on the bias curves of Figure 1 are very
small.  We can see this in several ways.  There is little fluctuation
along each dashed or solid curve, even though neighboring points refer
to disjoint $\nu$ ranges with no halos in common, and so should have
uncorrelated sampling uncertainties. Then we have estimated
uncertainties for each dashed and solid curve by measuring the {\it
rms} scatter in $b$ among 100 bootstrap resamplings of the halo
subsamples corresponding to the central and the largest values of
$\nu$ plotted.  In every case the scatter at corresponding points on
the solid and dashed curves is consistent within the estimation
uncertainties. We thus average these values in quadrature, listing the
results in parentheses in each panel to represent ``typical'' and
``maximal'' uncertainties. Typical uncertainties in $b$ are all
between 2 and 3\%, while maximal uncertainties are around 5\%. At the
small $\nu$ end of each curve bootstrap errors are always below 1\%.
Finally, and for us most convincingly, the overlap between the curves
for different redshifts is excellent in almost all cases.  Panel (f)
shows that the relevant mass limits differ by one or two orders of
magnitude, so that the samples involved are almost disjoint. In
addition, properties such as formation time or concentration are
independently evaluated at the different redshifts so that there is
almost no correlation between the particular objects that fall into
the 20\% tails.

\section{Conclusion}
In this Letter, we have used the very large Millennium Simulation to study
assembly bias, the fact that the large-scale clustering of haloes of given
mass depends on the details of how they were assembled. We have extended
earlier work by using numerical data from four different redshifts, by using
a less noisy estimator of clustering strength, and by studying bias effects as
a function of five physical properties of haloes in addition to their mass.

Assembly bias manifests itself in significantly and qualitatively different
ways for each of the five halo properties we have considered, although for all
of them our results are consistent with a dependence on mass and redshift
through the single parameter $\nu(M,z)$. The differences between our spin
results and those of Bett et al. (2007) suggest a dependence on the detailed
definition of spin which may parallel that on the detailed definition of
substructure fraction which we found above. Assembly bias varies with halo
properties at well above the 10\% level in all the cases we have studied. This
suggests that the large-scale clustering of galaxies cannot be used to infer
the amplitude of mass fluctuations to few percent accuracy without detailed
simulations of galaxy formation throughout large volumes.  Indeed, the
sensitivity of assembly bias to apparently minor details of halo formation
history may make it impossible, even with simulations, to achieve robust
results of the desired accuracy. Galaxies are complex objects and they may not
be suited to precision cosmology.

\section*{Acknowledgements}
We thank the referee for helpful comments. The authors are grateful to
the Virgo Consortium, and in particular to Volker Springel, for the
tremendous amount of work they invested to carry out the Millennium
Simulation and to make its results available for analysis. GL also
thanks Carlos Frenk, Shaun Cole, Adrian Jenkins and Cedric Lacey for
stimulating discussions. We are grateful to Eric Hayashi for checking
our numerical estimates of cross-correlations. Databases containing
halo/subhalo data at all times, as well halo formation trees and
galaxy properties for two independent galaxy formation simulations are
available at http://www.mpa-garching.mpg.de/millennium 
\label{lastpage}

\end{document}